\documentclass[a4paper,11pt]{article}
\usepackage{pos}
\usepackage{setspace}

\usepackage{comment}
\usepackage{cleveref}

\usepackage{graphicx}
\usepackage[caption=false]{subfig}	

\title{Absorption features in gamma-ray spectra of BL Lac objects}

\author*[a]{L. Foffano}
\author[a]{V. Vittorini}
\author[a,b,c,d]{M. Tavani}
\author[a]{E. Menegoni}

\emailAdd{luca.foffano@inaf.it}

\affiliation[a]{INAF - IAPS, via del Fosso del Cavaliere 100, I-00133 Roma (Italy)}
\affiliation[b]{Astronomia, Accademia Nazionale dei Lincei, via della Lungara 10, I-00165 Roma, Italy}
\affiliation[c]{Università ``Tor Vergata'', Dipartimento di Fisica, via della Ricerca Scientifica 1, I-00133 Roma, Italy}
\affiliation[d]{ Gran Sasso Science Institute, viale Francesco Crispi 7, I-67100 L’Aquila, Italy}

\abstract{
The production site of gamma rays in blazars is closely related to their interaction with the photon fields surrounding the active galactic nucleus. In this work we discuss an indirect method that may help to unveil the presence of ambient structures in BL Lacs through the analysis of their gamma-ray spectrum.
Passing through structures at different distances from the black hole, gamma rays interact with the corresponding photon fields via gamma-gamma pair production, producing absorption features in their spectral energy distribution. An interaction of the gamma-ray photons with a putative broad-line region may reduce the gamma-ray flux only if its production site were very close to the central engine. On the other hand, if jet photons interact with optical-UV seed photons produced by a pc-scale narrow-line region, the consequent gamma-gamma process may cause absorption features at a few hundreds GeV.
Sources with spectra reaching TeV energies, such as HBLs and EHBLs (extreme blazars), may represent exceptional probes to investigate this topic. In this regard, we discuss recent observations of sources which may show evidence of such absorption features in their gamma-ray spectra.
Finally, we discuss how sub-TeV absorption features in the spectra of BL Lacs may affect their broadband modeling, and eventually represent a powerful diagnostic tool to constrain the gamma-ray production site and the jet environment.
}

\FullConference{%
  7th Heidelberg International Symposium on High-Energy Gamma-Ray Astronomy (Gamma2022)\\
  4-8 July 2022\\
  Barcelona, Spain\\}

\begin{document}
\maketitle

Active galactic nuclei (AGNs) are among the most studied extragalactic objects in modern astrophysics. However, the determination of their precise structure, components, and emission mechanisms are still under debate. In \emph{blazars} - which are AGNs that emit relativistic jets pointing toward the line of sight of the observer - a standard method to investigate the presence of ambient photon fields is based on their optical spectra. 
This method can be easily applied in Flat Spectrum Radio Quasars (FSRQs), which are blazars with an optical spectrum rich in absorption and emission lines.  Conversely, BL~Lac objects are blazars where these emission lines are faint or not present at all, as the thermal components are overwhelmed by the non-thermal radiation of the jet. For this reason, this approach investigating the optical spectral lines is not adequate for these sources, and alternative methods need to be developed.

\section{Method}

In this work, we propose an alternative technique that may unveil the presence of large-scale structures through the analysis of the gamma-ray spectra of BL Lac objects. 
Specifically, we study the effects of an interaction between the relativistic jet of a blazar and the ambient photon fields produced by such large-scale structures surrounding the AGN \citep{Foffano2022}. 
Since this interaction has been systematically studied in FSRQs in \citep{costamante-blr}, in this work we will apply this method to another specific category of blazars, i.e. the BL Lac objects, where a complete discussion is still missing.

BL~Lac objects are sources with scarcely contaminated non-thermal emission of the jet that produces a spectral energy distribution (SED) deeply entering the gamma-ray region.
They are classified on the basis of their synchrotron peak frequency $\nu_{\text{peak}}^{\text{sync}}$ \citep{2010ApJ...716...30A}. In this paper, we will mainly consider high-peaked BL~Lac objects (HBL, $\nu_{\text{peak}}^{\text{sync}}$ between $10^{15}$ and $10^{17}$ Hz) and  extremely high-peaked objects (EHBLs or also \emph{extreme blazars}, defined when $\nu_{\text{peak}}^{\text{sync}} > 10^{17}$~Hz \citep{foffano-2019}).

\section{Photon-photon interaction and absorption process}
\label{sec:gammagammainteraction}

\begin{figure}
\centering
\includegraphics[width=0.75\textwidth]{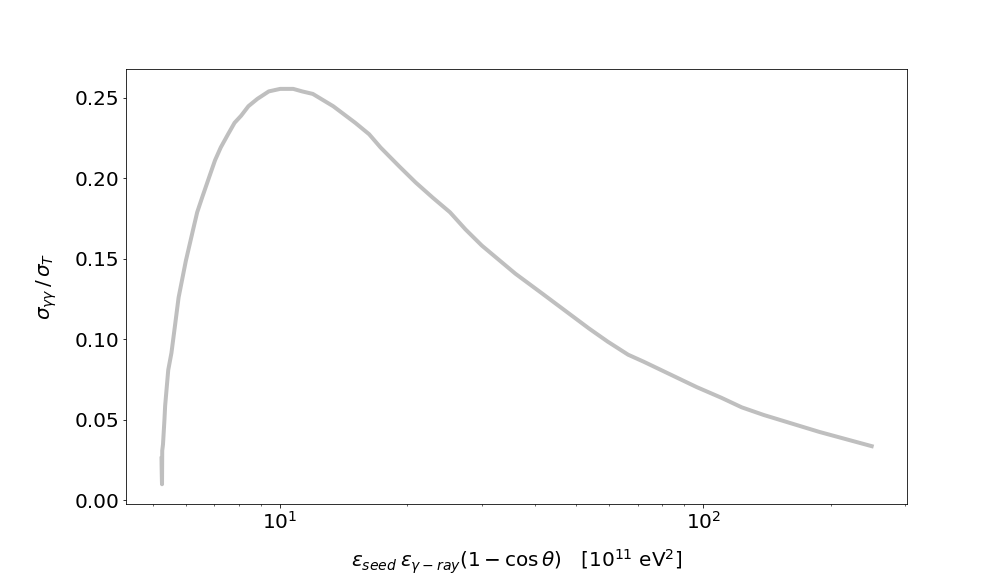}
\caption{Cross-section of the $\gamma\gamma$ interaction producing electron-positron pairs in units of the Thomson cross section $\sigma_T$ and as a function of the product of the interacting photon energies.}
\label{fig:gammagammacrosssection}
\end{figure}

Our study will assume the presence of a large-scale environment located on the trajectory of the relativistic jet of the BL Lac object. 
In this scenario, highly energetic photons of the relativistic jet $\gamma_{\text{jet}}$ would interact  via photon-photon $\gamma\gamma$ interaction process with a target photon field $\gamma_{\text{target}}$ emitted by the large-scale structure, producing an electron-positron pair
\[
\gamma_{\text{jet}} + \gamma_{\text{target}} \to e^+ + e^- \: .
\]
The cross-section of this interaction, represented in \Cref{fig:gammagammacrosssection}, is characterized by typical features, such as 1) a precise energy threshold, above which the production of pairs takes place, 2) a maximum value at twice the energy threshold, and 3) a decreasing trend as the energy of the incoming photons increases \citep[for more details see][]{aharonian2004book}.

In our scenario, the intrinsic emission of the relativistic jet $I_{\text{in}}$ would be reduced by this interaction, following 
\begin{equation}
I_{\text{out}}  = I_{\text{in}} \: e^{-\tau_{\gamma\gamma}} \; ,
\label{eq:absorption-flux}
\end{equation}
where  $I_{\text{out}}$ is the observed flux and $\tau_{\gamma\gamma}$ is the absorption factor of the $\gamma\gamma$ interaction. 
The latter can be expressed as a function of the cross-section $\sigma_{\gamma\gamma}$, of the size of the interacting region $R$, and of its photon density $n_{\text{seed}}$.
In this work, we will mainly focus on a mono-energetic, isotropic, and uniform seed photon field, adopting:
\begin{equation}
\tau_{\gamma\gamma} = \; n_{\text{seed}} \cdot R \cdot \sigma_{\gamma\gamma}(E) \; \simeq \;0.68 \cdot  n_{\text{seed,4}} \cdot (R/100\text{ pc})\; ,
\label{eq:tau-definition}
\end{equation}
where we adopt the notation $n_{\text{seed}} = 10^4 \: n_{\text{seed,4}}$. Specifically, we will define the target photon column density $K_{\text{seed}}$ as 
\[
K_{\text{seed}} = n_{{\text{seed}}} \cdot R \; ,
\]
with notation expressed in $\frac{pc}{cm^3}$ to explicit typical values of size and photon density.

\section{Expected observational effects}

The characterization of large-scale structures surrounding blazar jets is also correlated with the production site of the most powerful radiation emitted by blazars, i.e. the gamma rays. If gamma rays are emitted before a dense target photon field, their observed flux will be severely affected (or even completely suppressed) by the $\gamma\gamma$ interaction following \Cref{eq:absorption-flux}.
Depending on the physical parameters characterizing the interacting region and interpreted mainly by the target photon column density $K_{\text{seed}}$, the absorption factor defined in \Cref{eq:tau-definition} can be significantly different.
In the case of AGN photon fields - that typically have energies ranging from infrared to optical and UV - such $\gamma\gamma$ interactions would produce the strongest effects in gamma rays \citep{Foffano2022}.

\begin{figure}
\centering
\includegraphics[width=0.6\textwidth]{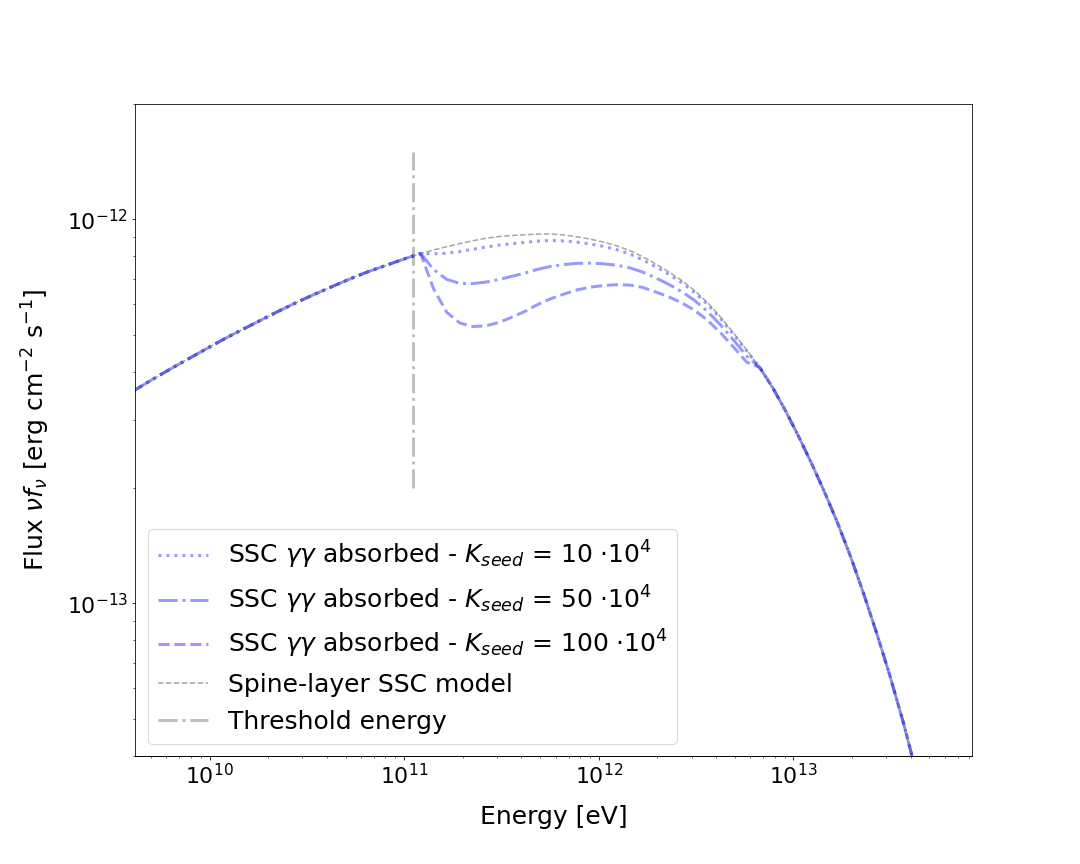}
\caption{Simulated observed gamma-ray spectrum of a typical \emph{extreme blazar} affected by $\gamma\gamma$ absorption resulting from interaction of photons of the blazar jet with seed photons of energy $\epsilon_{\text{seed}}\simeq4$~eV. We show several hypothetical curves related to different photon column densities $K_{\text{seed}}$ of the seed photon fields. The spectrum is already de-absorbed by the interaction with the EBL.}
\label{fig:generic_gammaray_spectrum_with_NLR_feature}
\end{figure}

An example is reported in \Cref{fig:generic_gammaray_spectrum_with_NLR_feature}, where we show how the observed spectrum of a hypothetical highly energetic BL Lac object (e.g. an \emph{extreme blazar}) would be affected by the $\gamma\gamma$ absorption process.  The spectrum would be modified depending on the energy of the jet photons, suffering a sharp reduction of the flux at the energy threshold of the reaction, and then a gradual reduction of this absorption effect.
The higher is $K_{\text{seed}}$, the deeper is the absorption feature in the gamma-ray spectrum. 
At larger energies of the seed photons correspond a lower threshold of the $\gamma\gamma$ interaction and a shift of the absorption feature to lower energies (and \emph{vice versa}).

\section{Comparison with real data}

\begin{figure}
\centering
\includegraphics[width=0.6\textwidth]{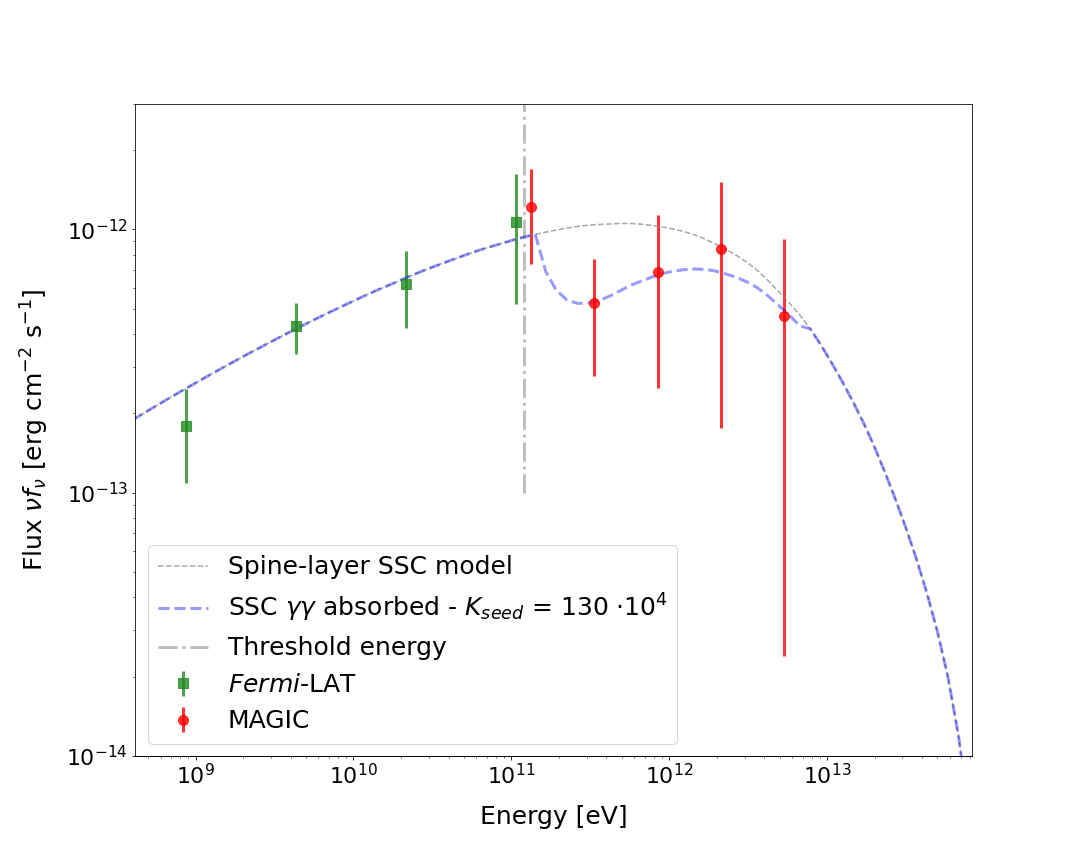}
\caption{Gamma-ray data (\emph{Fermi}-LAT data in green squares and MAGIC data in red circles) of the MWL SED of PGC~2402248 \citep[adapted from][]{MAGIC_paper_pgc}. We show also a theoretical model (grey dotted curve) and the additional $\gamma\gamma$~absorption resulting from the interaction with seed photons of energy $\epsilon_{\text{seed}}\simeq 4$ eV (blue dashed line), happening above the threshold energy (vertical line).}
\label{fig:pgc_gammaray_data}
\end{figure}

We have first applied this method to a BL Lac object recently discovered at TeV gamma-ray energies. This source, named PGC~2402248 (or 2WHSP~J073326.7+515354, $z = 0.065$), has been published by the MAGIC Collaboration in 2019 \citep{MAGIC_paper_pgc}. In \Cref{fig:pgc_gammaray_data} we show the available dataset of the intrinsic spectrum at gamma-ray energies extending up to 5~TeV (already \mbox{de-absorbed} by the interaction with the extragalactic background light, EBL, see e.g. \citep{Hauser:2001, Franceschini17}), which shows a continuously increasing flux up to about 100~GeV. At that energy, \emph{Fermi}-LAT and MAGIC data are well compatible, but at slightly higher energies of $\sim 250$~GeV the flux decreases to $\text{Flux}_{250\text{~GeV}} \simeq \frac{1}{2} \text{Flux}_{100\text{~GeV}}$. Then, the flux grows up again gradually to rejoin the expected emission at about a few TeV.

Assuming that the absorption feature in the spectrum of the source is due to $\gamma\gamma$ interaction as described in Section~\ref{sec:gammagammainteraction}, we can estimate the physical parameters of the large-scale interacting region. We note that the maximum of the interaction happens at about 250~GeV, and then we assume that the target photon field has energy in the optical-UV of about 4~eV. This value is compatible with the strongest emission lines of the spectrum of  well-studied narrow-line radio galaxies \citep[e.g.][]{osterbrock} and also with the faint emission lines emerging in the optical spectrum of our source PGC~2402248 in \citep{pgc_optical_spectrum}.
For continuity of the spectral emission in the gamma-ray spectrum, we assume that the input flux before entering the absorption region at 250~GeV was $I_{\text{in}} \equiv \text{Flux}_{100\text{GeV}}$ and that the observed flux is $I_{\text{out}} \equiv \frac{1}{2} \text{Flux}_{100\text{~GeV}}$. From  \Cref{eq:absorption-flux}, this implies that the maximum absorption factor is $\tau_{\gamma\gamma} = 0.70 \pm 0.36$.
Then, from \Cref{eq:tau-definition}, we extract the total photon column density that may produce this absorption factor as $K_{\text{seed,max}} = (104\pm37) \cdot 10^4 \; \frac{pc}{cm^3}$.
The corresponding parameter space of the properties of the interacting region defining $K_{\text{seed}}$ - size and photon density - is compatible with the properties of known large-scale structures of AGNs, such as NLR and extended NLR.

Concerning the origin of the required target photon density, similarly to the most common models of BLR radiation, the NLR may be photoionized by the accretion disk. However, the distance between the photon source and the target large-scale structure implies that is difficult to obtain the required photon density with the usual luminosity of accretion disks.
As an alternative scenario, the enlightenment of the NLR may come from the radiation emitted by the jet itself. The synchrotron radiation emitted before or while passing through the NLR  would radiate photons and illuminate its clouds, that in turn may re-emit and increase locally the seed photon density. Such a target photon density would easily reach the values for the $\gamma\gamma$ process with the gamma~rays of the jet to happen, and produce the observed reduction of the gamma-ray photon flux.
This scenario, as demonstrated in \citep{vittorini2014}, provides higher photon densities thanks to the boost given by the moving plasmoid and to its lower distance from the reflecting clouds.

\section{Conclusions}

In this work \cite{Foffano2022}, we study the effects of an interaction between a putative narrow-line region and the relativistic jet of a BL Lac object. In this scenario, highly energetic radiation of the jet  would interact via  $\gamma\gamma$ pair-production process with low-energy seed photons emitted by AGN structures close to the jet trajectory. Such a process would mostly affect the gamma-ray spectrum at a few hundreds GeV, and produce detectable absorption features in the sub-TeV spectrum. 

Experimentally, this method may be studied with an inverse procedure, i.e. by verifying the presence of absorption features in the gamma-ray spectra of BL Lac objects. Such features may indirectly unveil the presence of ambient structures that can not be identified with standard methods using their optical spectra. 

This absorption effect may be detected only on sources with favourable spectral properties, and that 1) are well detected in VHE gamma rays, 2) show hard spectrum extending up to several hundreds GeV, 3) show a clean spectral shape (without contamination by other spectral features) in that band, and 4) show a relatively stable flux at those energies.

Interestingly, the most promising sources where this effect could be easily detectable are HBLs and EHBLs (\emph{extreme blazars}), with their emission reaching TeV energies. 
In this regard, we discussed the case of the extreme blazar PGC~2402248 (also named 2WHSP~J073326.7+515354) recently detected at TeV energies by the MAGIC Collaboration, which shows an evidence of spectral absorption that may be due to a $\gamma\gamma$ process. In this picture, the seed photons are compatible with those emitted by a narrow-line region located along the jet path, which is illuminated by the jet itself and responsible for the partial absorption of sub-TeV gamma rays.

A systematic analysis of well-sampled gamma-ray spectra of BL Lac objects will be particularly important to verify this scenario, and will be presented in a forthcoming paper.\\

\noindent
\textbf{Acknowledgements} Research carried out through partial support of the ASI/INAF AGILE contract I/028/12/05.

\setstretch{0.5}
\bibliography{bib_gamma}
\bibliographystyle{aasjournal}

\end{document}